\documentclass{cernepprep} 
\usepackage{cernunits,cernchemsym,heppennames2}
\usepackage{cite,fancyvrb}

\usepackage{graphicx}
\usepackage{dcolumn}
\usepackage{bm}
\usepackage[utf8]{inputenc}
\usepackage[T1]{fontenc}
\usepackage{newtx}
\usepackage{etoolbox}
\usepackage{amsmath}
\usepackage{import}
\usepackage{enumitem}
\usepackage{subfigure}
\usepackage[subfigure]{tocloft}
\usepackage{multirow}
\usepackage{overpic}
\usepackage{tabularx}
\usepackage{setspace}
\usepackage{anyfontsize}
\usepackage[T1]{fontenc}
\usepackage[version=4]{mhchem}
\usepackage{gensymb}
\usepackage{hyperref}

\begin{document}

\begin{titlepage}
\EPnumber{2025-259}
\EPdate{\today}
\DEFCOL{CDS-Library}

\title{Optimizing Antihydrogen Production via Slow Plasma Merging}

E.~D.~Hunter$^{1,2}$\footnote{Electronic mail: eric.david.hunter@cern.ch},
M.~Bumbar$^{1,2,3}$,
C.~Amsler$^{4}$\footnote{Current Address: Marietta Blau Institute for Particle Physics, Austrian Academy of Sciences, 1010 Vienna, Austria},
M.~N.~Bayo$^{5,6}$,
H.~Breuker$^{7}$, 
M. Cerwenka$^{4,3}$\textsuperscript{\textdagger},
G. Costantini$^{8,9}$,
R. Ferragut$^{5,6}$,
M. Giammarchi$^{4}$,
A. Gligorova$^{4}$\footnote{Current Address: Faculty of Physics, University of Vienna, Boltzmanngasse 5, 1090 Vienna, Austria}, 
G. Gosta$^{8,9}$,
M.~Hori$^{2,10}$,
C.~Killian$^{4}$\textsuperscript{\textdagger},
V.~Kraxberger$^{4,3}$\textsuperscript{\textdagger},
N.~Kuroda$^{11}$,
A.~Lanz$^{4,3}$\footnote{Current Address: University College London, London WC1E 6BT, United Kingdom}, 
M.~Leali$^{8,9}$,
G.~Maero$^{12,6}$,
C.~Mal\-bru\-not$^{13}$
V.~Mascagna$^{8,9}$,
Y.~Matsuda$^{11}$,
S.~Migliorati$^{8,9}$,
D.~J.~Murtagh$^{4}$\textsuperscript{\textdagger},
M.~Romé$^{12,6}$,
R.~E.~Sheldon$^{4}$\textsuperscript{\textdagger},
M.~C.~Simon$^{4}$\textsuperscript{\textdagger},
M.~Tajima$^{14,2}$,
V. Toso$^{6,12}$,
S.~Ulmer$^{7,15}$,
L.~Venturel\-li$^{8,9}$,
A.~Weiser$^{4,3}$\textsuperscript{\textdagger},
E.~Wid\-mann$^{4}$\textsuperscript{\textdagger}\\

\centering{(The ASACUSA-Cusp Collaboration)}\\[20pt]

$^1$CERN, 1211 Geneva 23, Switzerland,
$^2$Imperial College London, London SW7 2BW, United Kingdom,
$^3$Vienna Doctoral School in Physics, University of Vienna, 1090 Vienna, Austria,
$^4$Stefan Meyer Institute for Subatomic Physics, Austrian Academy of Sciences, 1010 Vienna, Austria,
$^5$L-NESS and Department of Physics, Politecnico di Milano, 22100 Como, Italy,
$^6$INFN sez. Milano, 20133 Milan, Italy,
$^7$Ulmer Fundamental Symmetries Laboratory, RIKEN, 351-0198 Saitama, Japan,
$^8$Diparti\-mento di Ingegneria dell'In\-formazione, Universit\`a degli Studi di Brescia, 25123 Brescia, Italy,
$^9$INFN sez. Pavia, 27100 Pavia, Italy,
$^{10}$Max-Planck-Insitut f\"ur Quantenoptik, D85748 Garching, Germany,
$^{11}$Institute of Physics, Graduate School of Arts and Sciences, University of Tokyo, 153-8902 Tokyo, Japan
$^{12}$Dipartimento di Fisica, Università degli Studi di Milano, 20133 Milan, Italy
$^{13}$TRIUMF, Vancouver BC V6T 2A3, Canada
$^{14}$Japan Synchrotron Radiation Research Institute, 1-1-1 Kouto, Sayo-cho, Sayo-gun, Hyogo 679-5198, Japan
$^{15}$Insitut f\"ur Experimentalphysik, Heinrich Heine Universit\"at, D\"usseldorf, Germany

\date{\today}

\begin{abstract}
We measure the time-dependent temperature and density distribution of antiprotons and positrons while slowly combining them to make antihydrogen atoms in a nested Penning-Malmberg trap. The total antihydrogen yield and the number of atoms escaping the trap as a beam are greatest when the positron temperature is lowest and when antiprotons enter the positron plasma at the smallest radius. We control these parameters by changing the rate at which we lower the electrostatic barrier between the antiproton and positron plasmas and by heating the positrons. With the optimal settings, we produce $2.3\times 10^6$ antihydrogen atoms per $15$-minute run, surpassing the previous state of the art\textemdash $3.1\times 10^4$ atoms in $4$ minutes\textemdash by a factor of $20$.
\end{abstract}

\end{titlepage}
\section{Introduction}
\label{sec:intro}
The observable universe seems to contain very little antimatter: mostly unstable decay products in cosmic ray showers, beta-decay positrons, and high energy cosmic rays \cite{aguilar_2025_antiprotons}. In the lab, we can trap a small amount\textemdash $10^7$ antiprotons ($\overline{\mathrm{p}}$) or $10^9$ positrons ($\mathrm{e}^+$)\textemdash as a nonneutral plasma \cite{kuroda_2012_development, blumer_2022_positron}. The plasma is confined by electric and magnetic fields in a Penning-Malmberg trap \cite{malmberg_1975_properties}. Four experiments at the CERN Antiproton Decelerator (AD) \cite{maury_1997_antiproton} presently use such plasmas to produce antihydrogen ($\overline{\mathrm{H}}$) \cite{hori_2013_physics}. $\overline{\mathrm{H}}$ is neutral \cite{amole_2014_an}, stable \cite{andresen_2011_confinement}, and can be compared with hydrogen spectroscopically \cite{bertsche_2015_physics} and gravitationally \cite{anderson_2023_observation}. Such comparisons may help to explain the observed excess of matter over antimatter \cite{sakharov_1967_violation}.

Experiments so far use a small fraction (${<}10^{-3}$) of the $\overline{\mathrm{H}}$ that they produce. Only the coldest atoms, with kinetic energy equivalent to a temperature $T<0.5\,\mathrm{K}$, can be trapped by the tesla-scale magnetic bottles used by the ALPHA and ATRAP collaborations \cite{andresen_2010_trapped,gabrielse_2012_trapped}. $T$ can be higher in beam experiments. The GBAR collaboration plans to create $\overline{\mathrm{H}}^+$ ions at a mean kinetic energy of $6\,\mathrm{keV}$ ($T\approx 7\times 10^7\,\mathrm{K}$). Our collaboration (ASACUSA) plans to measure the ground-state hyperfine structure in a beam of $\overline{\mathrm{H}}$ \cite{widmann_2004_measurement}. Our spectroscopy apparatus can analyze beams of velocity $\upsilon \lesssim 1500\,\mathrm{m/s}$, corresponding to $T\approx 50\,\mathrm{K}$ \cite{malbrunot_2019_hydrogen}. However, the short time, about $1\,\mathrm{ms}$ between $\overline{\mathrm{H}}$ formation and analysis, means that most atoms in the beam are not in the ground state \cite{robicheaux_2004_simulations,radics_2014_scaling,kolbinger_2021_measurement}. Moreover, most atoms are not 'beam-like' (velocity vector within a few degrees of the trap axis). This motivates our work to increase the number of $\overline{\mathrm{H}}$ atoms, to measure their properties shortly after creation, and to control those properties.

In this article, we study $\overline{\mathrm{H}}$ production and beam formation using the method known as ``slow-merge mixing,'' whereby $\overline{\mathrm{p}}$ and $\mathrm{e^+}$ plasmas are combined by adjusting the electrostatic potentials that confine them \cite{gabrielse_1989_possible}. We use an order of magnitude more particles of each species to produce 70 times more $\overline{\mathrm{H}}$ per run than the previous state of the art \cite{ahmadi_2017_antihydrogen}, with $70$ to $80\%$ of the input $\overline{\mathrm{p}}$ forming stable $\overline{\mathrm{H}}$. (A ``run'' is a 15-minute sequence of operations to prepare the plasmas, make $\overline{\mathrm{H}}$, and diagnose the remaining particles.) We also observe a beam of $\overline{\mathrm{H}}$ atoms leaving the trap, which we characterize in another article \cite{hunter_2025_measured}. Here, we focus on the plasma properties, describing the plasma preparation (and detectors) (Section~\ref{sec:app}); time-dependent changes in the plasma density profile, $T$,  and $\overline{\mathrm{H}}$ production (Section~\ref{sec:time}); the slow extraction of $\overline{\mathrm{p}}$ from the plasma (Section~\ref{sec:slowext}); and the effect of the mixing duration and $T$ of the $\mathrm{e^+}$ (Section~\ref{sec:rate}). We summarize our findings (Section~\ref{sec:disc}) and discuss the analysis of plasma properties (Appendix~\ref{sec:appx_plasma}).

\section{Experiment}
\label{sec:app}

\begin{figure}
    \centering
    \includegraphics[width=0.95\linewidth]{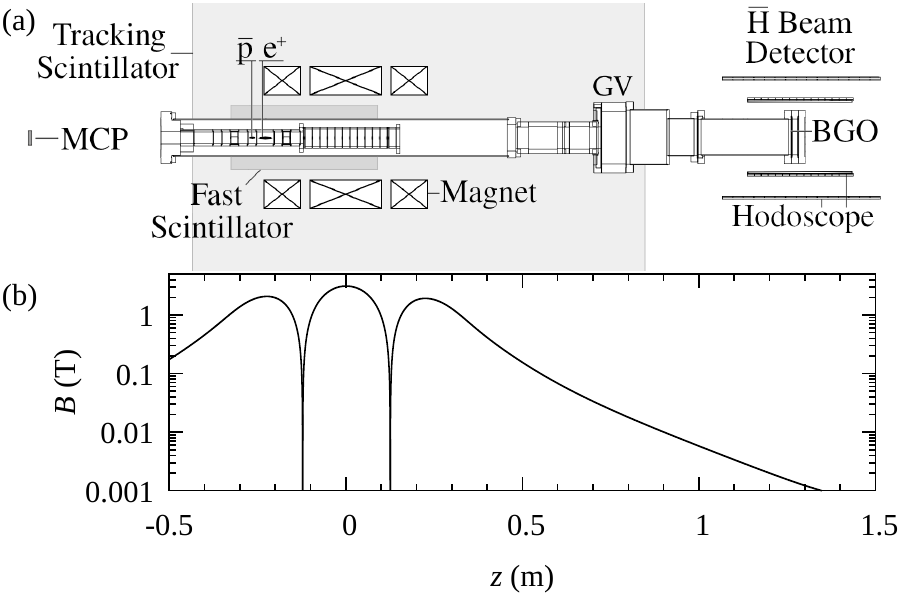}
    \caption{Simplified cross section of the experiment and calculated magnetic field $B$ along the axis of symmetry. Drawing in (a) is scaled to match the axial position $z$ in (b). The downstream direction corresponds to increasing $z$. Abbreviations: Scintillating bar arrays (Tracking Scintillator), microchannel-plate phosphor screen (MCP), inner bore scintillator plates (Fast Scintillator), Cusp magnet coils (Magnet), gate valve (GV), and $\overline{\mathrm{H}}$ Beam Detector consisting of bismuth germanium oxide crystal (BGO) and scintillating bar layers (Hodoscope). We use the symbols $\overline{\mathrm{p}}$ and $\mathrm{e^+}$ for the antiproton and positron plasmas. Microwave meshes at $z=-0.52$ and $+0.48\,\mathrm{m}$ and field ionizers at $z=-0.12$ and $+0.12\,\mathrm{m}$ are not shown.}
    \label{fig:appa}
\end{figure}

\begin{figure}
    \centering
    \includegraphics[width=0.95\linewidth]{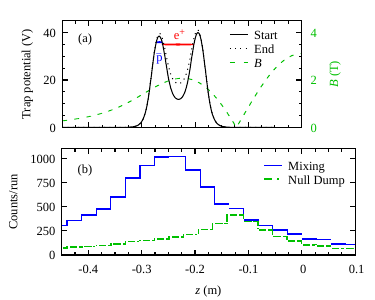}
    \caption{Trapping potentials and annihilation vertex positions. (a) The electric potential at $r=0$ generated by the biased electrodes at the start and end of mixing, for the protocol used in Section~\ref{sec:time}. $B$ is given again for reference. Length and space charge $\phi$ of the $\overline{\mathrm{p}}$ and $\mathrm{e^+}$ plasmas are suggested by the blue and red ellipses. (b) Vertices found using the Tracking Scintillator during $\overline{\mathrm{H}}$ production (``Mixing'') and slow release downstream (``Null dump'').}
    \label{fig:brescia}
\end{figure}

The Cusp trap \cite{enomoto_2010_synthesis}, shown in Fig.~\ref{fig:appa}(a), is where we combine $\overline{\mathrm{p}}$ and $\mathrm{e^+}$ plasmas to make $\overline{\mathrm{H}}$. The trap magnetic field $B$, shown in panel (b), confines the $\mathrm{e^+}$ and $\overline{\mathrm{p}}$ radially. The strong gradients in $B$ can focus low-field-seeking atoms \cite{nagata_2014_a,nagata_2015_the} and do not degrade standard plasma manipulations \cite{hunter_2023_sdr}. We apply time-dependent voltages on hollow cylindrical electrodes to confine and manipulate particles axially. Figure~\ref{fig:brescia}(a) shows the trapping fields at the start and end of $\overline{\mathrm{H}}$ production. Trap electrodes have inner diameter $34\,\mathrm{mm}$ ($50\,\mathrm{mm}$ for electrodes outside the trapping region). Most trap electrodes have in-trap low-pass filters with time constant $RC \approx 0.01\,\mathrm{ms}$. Three of the electrodes, $0.05$ to $0.25\,\mathrm{m}$ away from the middle of the mixing region, do not have filters and are used for plasma compression, heating, pulsed extraction, and transfer. The entrance and exit to the trap are screened by $79\%$ transparent copper meshes to reflect microwave radiation from warm objects outside the $6\,\mathrm{K}$ region ($z=-0.5$ to $0.5\,\mathrm{m}$ in Fig.~\ref{fig:appa}), which would heat the $\mathrm{e^+}$ \cite{amsler_2022_reducing}. The $\overline{\mathrm{p}}$ lifetime exceeds $6000\,\mathrm{s}$, implying that the pressure is lower than $10^{-13}\,\mathrm{mbar}$ \cite{sellner_2017_improved}.

Figure~\ref{fig:appa}(a) shows the locations of our detectors. We use a single-stage microchannel plate-phosphor screen detector (MCP), with a silicon photomultiplier (SiPM) for diagnosing the plasma temperature $T$ and for counting $\overline{\mathrm{p}}$ as in Refs.~\cite{eggleston_1992_parallel, hunter_2020_plasma, frederiksen_2022_counting}. When counting $\overline{\mathrm{p}}$, we release particles upstream toward the MCP at $z=-0.9\,\mathrm{m}$, reducing the confining potential slowly enough to resolve each hit \cite{cripe_2024_summer}. The detection efficiency is close to $100\%$ (see Appendix~\ref{sec:appx_plasma}). A CMOS camera records the phosphor screen images, which contain information on the radial distribution of plasma particles \cite{andresen_2009_antiproton}. Inside the magnet bore, 8 long scintillator plates (Fast Scintillator) record the rate of $\overline{\mathrm{p}}$ annihilation in the trap \cite{cripe_2024_summer}. Appendix~\ref{sec:appx_plasma} describes our method of calibrating the Fast Scintillator. Outside the magnet, 8 arrays of scintillating bars (Tracking Scintillator) track pions generated when $\overline{\mathrm{p}}$ annihilate. The distribution of reconstructed annihilations is smeared out by scattering of the pions in the Cusp magnet, yielding a peak position accuracy of about $0.03\,\mathrm{m}$ in the axial coordinate $z$ \cite{costantini_2023_the}. Figure~\ref{fig:brescia}(b) shows the output from the Tracking Scintillator, averaged over 81 runs. Most events are close to the $\mathrm{e^+}$ plasma during mixing and close to the minimum of $B$ when $\overline{\mathrm{p}}$ are dumped downstream. The rate limit of the Tracking Scintillator is about $1\,\mathrm{kHz}$, which is too slow for the event rate to be proportional to the true annihilation rate (for which we use the Fast Scintillator). To detect $\overline{\mathrm{H}}$ that leave the trap downstream, we use a bismuth germanium oxide crystal (BGO) and two hodoscope layers in 3-coincidence counting mode, as described in \cite{nagata_2017_development,hunter_2025_measured}. The BGO-hodoscope combination is called ``$\overline{\mathrm{H}}$ beam detector'' and provides the signal for ASACUSA's spectroscopy program. An average rate of background events\textemdash mostly from pions created when $\overline{\mathrm{H}}$ annihilates on the trap electrodes\textemdash is subtracted using data from runs with the gate valve (GV) closed. 

One bunch of $100\,\mathrm{keV}$ $\overline{\mathrm{p}}$ from ELENA accelerates to $120\,\mathrm{keV}$ in ASACUSA's $20\,\mathrm{kV}$ pulsed drift tube before hitting a pair of roughly $700$-nm-thick degrader foils \cite{amsler_2024_injection}. We capture $25\%$ of the bunch, or about $3\times 10^6$ $\overline{\mathrm{p}}$ per AD cycle. The $\overline{\mathrm{p}}$ from multiple AD cycles can be collected in our catching trap \cite{kuroda_2012_development}. We take either 3 bunches (Sections~\ref{sec:time},\ref{sec:slowext}) or 1 bunch (Section~\ref{sec:rate}) before compressing the $\overline{\mathrm{p}}$ and transferring them to the Cusp trap, where they are electron-cooled and compressed again. Then we remove the electrons \cite{andresen_2008_compression}, leaving a plasma of ${>}99\%$ $\overline{\mathrm{p}}$ with the properties given in Table~\ref{tab:initial}.

{\renewcommand{\arraystretch}{1.2}
\setlength{\tabcolsep}{0.5em} 
\begin{table}
  \begin{center}
  \begin{tabular}{c|c|c|c|c|c|c}
      Species & Section & Type & $N\,\mathrm{(10^6)}$  & $T\,\mathrm{(K)}$ & $r_0\,\mathrm{(mm)}$ & $\lambda_\mathrm{D}\,\mathrm{(mm)}$ \\
        \hline
        $\mathrm{e}^+$          & III  & exp.      &   $101\pm 2$   &  $50\pm 20$        &   $1.50\pm 0.01$  & - \\
                 &        &   num. & $100\pm 10$   &  $50\pm 20$        &   $1.4\pm 0.2$  & $0.036\pm 0.009$ \\
                 & V  & exp.        &   $64\pm 1$     &  $50\pm 20$        &   $1.26\pm 0.02$ & - \\
                  &        &  num. & $64\pm 9$   &  $50\pm 20$        &   $1.3\pm 0.2$  & $0.041\pm 0.009$ \\
                  \hline
        $\overline{\mathrm{p}}$ & III & exp.  &   $2.89\pm 0.04$   &   $1700\pm 200$    &   $1.14\pm 0.04$  & - \\
         &      &   num. & $2.9\pm 0.3$   &   $1700\pm 200$    &   $1.13\pm 0.07$  & $0.32\pm 0.02$ \\
         & V  & exp.       &   $1.7\pm 0.1$   &  $700\pm 100$     &   $1.07\pm 0.06$ & - \\
         &  & num. & $1.7\pm 0.2$   &  $700\pm 100$     &   $1.05\pm 0.06$ & $0.27\pm 0.03$  \\
  \end{tabular}
  \caption{Plasma properties: number of particles ($N$), temperature ($T$), radius ($r_0$), and Debye length ($\lambda_\mathrm{D}$). The roman numerals refer to the corresponding section in this article. The $\mathrm{e^+}$ plasma properties are averaged over the $50$ to $60\,\mathrm{s}$ mixing duration, while the $\overline{\mathrm{p}}$ ones are given only for the start of mixing, as they change significantly in time. Results from a numerical solver appear below each line of measured values. Measured values (exp.) are accompanied by statistical uncertainties, and solver values (num.) by systematic uncertainties. See Appendix~\ref{sec:appx_plasma}.}
  \label{tab:initial}
  \end{center}
\end{table}
}

We trap beta-decay $\mathrm{e^+}$ from a \ce{^{22}Na} source using a First Point Scientific system~\cite{lanz_2023_upgrade}. Every $1.05\,\mathrm{s}$ we move the $\mathrm{e^+}$ to another trap at lower pressure, repeating until about $10^7\,\mathrm{e^+}$ are in the low-pressure trap. We successively collect and transfer several such bunches from the low-pressure trap to the Cusp trap. We purify the $\mathrm{e^+}$ plasma ($99.98\%\,\mathrm{e^+}$) to reduce the plasma expansion heating caused by positive ion contamination \cite{hunter_2025_best} and compress it using the SDREVC technique \cite{ahmadi_2018_enhanced,hunter_2023_sdr}, which ensures $1\%$-level reproducibility of the plasma radius $r_0$ and number of particles $N$. Note that
the systematic uncertainty in estimating \emph{in situ} plasma properties (even rows of Table~\ref{tab:initial}) from measured values (odd rows) is greater than $1\%$. See Appendix~\ref{sec:appx_plasma} for more details.

In mixing experiments, we ramp the electrode voltages so that the electrostatic potential morphs linearly from ``Start'' to ``End,'' as shown in Fig.~\ref{fig:brescia}(a). Then we ramp the potentials back to ``Start'' and separate the remaining particles. In most cases, we count the leftover $\overline{\mathrm{p}}$ by releasing them downstream (``Null dump'')\textemdash they hit the wall and annihilate near the field null at $z=-0.12\,\mathrm{m}$\textemdash and we keep most of the $\mathrm{e^+}$ for the next run, releasing $10\%$ of them upstream to measure $T$. Otherwise we dump each plasma as a whole to the MCP, one after the other, as quickly as possible so that the instabilities described in Section~\ref{sec:slowext} do not distort the density profiles, which we will present first in the following section. 

\section{Trends during mixing}
\label{sec:time}
\begin{figure}
    \centering
    \includegraphics[width=0.95\linewidth]{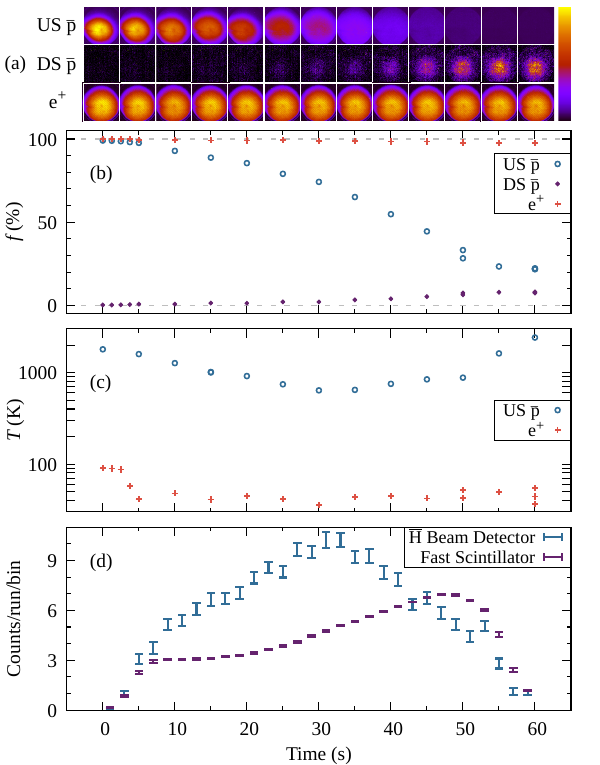}
    \caption{Plasma diagnosed partway through mixing: (a) images, (b) fraction remaining $f$, (c) plasma temperature $T$, (d) count rate at the $\overline{\mathrm{H}}$ Beam Detector and $1/1000$ times the count rate at the Fast Scintillator. In (a), the side-length of the image squares is $4\,\mathrm{cm}$, or about $5\,\mathrm{mm}$ in the trap (see Appendix~\ref{sec:appx_plasma}). The intensity scale is the same for all images in a given row. The color gradient corresponding to normalized intensity range $0$ to $1$ is given at the right of the image squares. The signals in (d) are the averages from 81 uninterrupted runs, with the standard deviation over these runs given as the error bars.}
    \label{fig:time}
\end{figure}

Figure~\ref{fig:time}(a) is a series of $z$-integrated density profiles obtained by interrupting the slow voltage ramp at different times during mixing and rapidly releasing plasma particles to the MCP. The plasma radius $r_0$ in Table~\ref{tab:initial} is derived from the profiles recorded at the start ($t=0$). During mixing, the $\overline{\mathrm{p}}$ in the upstream (US) well expand, and some are captured at low radius in the downstream (DS) well. The $\mathrm{e^+}$ plasma does not expand measurably in this time. 

Panel~(b) complements (a) by quantifying the loss of $\overline{\mathrm{p}}$ from the US well, the accumulation of $\overline{\mathrm{p}}$ in the DS well, and the roughly constant number of $\mathrm{e^+}$. The fraction $f$ is the number of particles found in the corresponding well when mixing is interrupted, divided by the number of particles in the trap at the start of mixing. For $\overline{\mathrm{p}}$ we calculate $f$ using Fast Scintillator data and for $\mathrm{e^+}$ we use their space charge potential $\phi_\mathrm{e}$ (see below). The drop in $f$ causes the $\overline{\mathrm{p}}$ plasma to expand \cite{andresen_2010_evaporative}, but not enough to account for the observed effect in panel (a). For $\overline{\mathrm{p}}$, the nested well has the form of a ``squeeze potential,'' which is known to enhance wave damping, transport, and heating \cite{kabantsev_2001_trappedparticle,dubin_2017_superbanana,anderegg_2019_plasma}.

Panel~\ref{fig:time}(c) shows $T$, which is measured with the MCP and SiPM as in Ref.~\cite{hunter_2020_plasma}.  We measure $T$ $0.2\,\mathrm{s}$ after the end of the mixing ramp so that the $\mathrm{e^+}$ do not have time to cool before being diagnosed. The $\overline{\mathrm{p}}$ temperature $T_\mathrm{p}$ falls during the first $30\,\mathrm{s}$ of mixing, then rises. The $\mathrm{e^+}$ temperature $T_\mathrm{e}$ jumps to $100\,\mathrm{K}$ for the first few seconds, then falls and stays at $(44\pm 5)\,\mathrm{K}$, which is also the steady state temperature with no $\overline{\mathrm{p}}$ in the trap. 

The count rate at the $\overline{\mathrm{H}}$ Beam Detector, shown in panel~(d), peaks when $T_\mathrm{p}$ reaches its minimum value. The Fast Scintillator count rate rises toward the end of mixing, indicating a higher rate of $\overline{\mathrm{p}}$ leaving the $\overline{\mathrm{p}}$ plasma. This data comes from uninterrupted mixing runs in which a field ionizer partially blocks weakly bound atoms\textemdash those with less than about $k_\mathrm{B}\times 200\,\mathrm{K}$ of binding energy, where $k_\mathrm{B}$ is Boltzmann's constant \cite{hunter_2025_measured}.

We estimate changes in $N$ as $\Delta N/N = \Delta f/f(t=0)$, where $N$ ($N_\mathrm{e}$ and $N_\mathrm{p}$ for $\mathrm{e^+}$ and $\overline{\mathrm{p}}$) is determined via separate measurements and verified numerically as described in Appendix~\ref{sec:appx_plasma}. We find that $|\Delta N_\mathrm{e}|=(2.9 \pm 0.3)\times 10^6\ \mathrm{e^+}$ and $|\Delta N_\mathrm{p}|=(2.3\pm 0.2)\times 10^6$ $\overline{\mathrm{p}}$ leave the plasma during mixing. The $\overline{\mathrm{H}}$ yield is $|\Delta N_\mathrm{p}|$ as bare $\overline{\mathrm{p}}$ cannot escape the confining potentials in the mixing region. $|\Delta N_\mathrm{e}|-|\Delta N_\mathrm{p}|=0.6\times 10^6\,\mathrm{e^+}$ pass through the $\overline{\mathrm{p}}$ plasma and hit the MCP instead of forming $\overline{\mathrm{H}}$. We will show in Section~\ref{sec:rate} that these escaping $\mathrm{e^+}$ reduce the $\overline{\mathrm{H}}$ yield by lowering $|\Delta N_\mathrm{p}|$.

\section{Analogy with slow extraction}
\label{sec:slowext}

We now present a $\overline{\mathrm{p}}$-only study that will help to understand some of the trends found in Section~\ref{sec:time}. The same $\overline{\mathrm{p}}$ plasma is used, with the difference that no $\mathrm{e^+}$ are loaded into the trap. We will use the term slow extraction for particles released gradually from the plasma, whether or not they escape the trap.

The radius $r$ at which $\overline{\mathrm{p}}$ enter the $\mathrm{e^+}$ plasma affects the velocity and binding energy of the $\overline{\mathrm{H}}$ formed \cite{jonsell_2009_simulation,jonsell_2019_formation}. We summarize these effects briefly. The electric field $\mathbf{E}$ of the $\mathrm{e^+}$ causes all charged particles in the plasma, including the $\overline{\mathrm{p}}$, to rotate with local velocity $\mathbf{E\times B}/B^2$. In a constant-density plasma, $\mathbf{E}$ is proportional to $r$. This rotation becomes a nearly radial velocity component when the $\overline{\mathrm{p}}$ is neutralized as $\overline{\mathrm{H}}$. Atoms formed at lower $r$ drift outward more slowly and take longer to reach the radial edge of the plasma. This increases the binding energy (on average) because collisions with $\mathrm{e^+}$ in the plasma tend to deexcite Rydberg $\overline{\mathrm{H}}$ \cite{robicheaux_2004_simulations,radics_2014_scaling}. 

No model predicts the distribution of $r$ for the $\overline{\mathrm{p}}$ in our case. At first, particles escape within $r<2\lambda_\mathrm{D}$, where $\lambda_\mathrm{D}=\sqrt{\epsilon_0 k_\mathrm{B}T/\rho e^2}$ is the Debye length, $\epsilon_0$ is the permittivity of free space, $\rho$ is the plasma density, and $e$ is the elementary charge \cite{danielson_2007_extraction,zhong_2024_shotnoiseinduced}. This follows from Gauss's law for a thermal distribution of charges in the limit that only a small fraction of the plasma has been extracted. It does not apply to the majority of the particles: by definition, a plasma is many $\lambda_\mathrm{D}$ thick, and even before the inner Debye cylinder is emptied, the hollowing of the plasma makes it unstable \cite{driscoll_1990_observation}. The diocotron instability radically changes the plasma density profile and is not \emph{a priori} predictable \cite{mason_2002_simulations}. 

We can directly measure $r$ for $\overline{\mathrm{p}}$ escaping the plasma in a situation analogous to mixing. Instead of releasing the $\overline{\mathrm{p}}$ into the $\mathrm{e^+}$, we release them towards the MCP. This does not include effects from $\phi_e$ or rebounding $\overline{\mathrm{p}}$, which are present during mixing. Nevertheless, we will use this analogy to identify controllable behavior which seems generalizable and important to the case of mixing. 

\begin{figure}
    \centering
    \includegraphics[width=\linewidth]{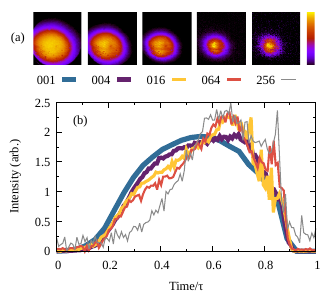}
    \caption{Slow extraction of $\overline{\mathrm{p}}$ to the MCP. (a) Transverse profiles integrated in time: each pixel is summed over all the frames of the corresponding video. Each image is then normalized to its own maximum. (b) Temporal profiles integrated in space: intensity is the sum of all pixel values in that frame. The intensity of each trace in (b) is normalized to its integral, and the time axis is rescaled such that 0 is the start of the extraction and 1 is the time $\tau$ (in seconds) given in the key under each image. The color gradient bar and the side-length of the image squares in (a) are the same as in Fig.~\ref{fig:time}.}
    \label{fig:ext}
\end{figure}

Figure~\ref{fig:ext} shows the spatial and temporal $\overline{\mathrm{p}}$ profiles measured during slow extraction. The $\overline{\mathrm{p}}$ are released to the MCP by removing the upstream barrier in a time $\tau$, which we vary between $1$ and $256\,\mathrm{s}$. A linear ramp of the voltage applied to the barrier electrode results in a nonlinear change in well depth vs.\ time. This means that the rate at which particles escape, shown in panel (b), would not be constant even if the particles were evenly distributed in energy and fixed at $r=0$. In fact, kinetic energy and $r$ also depend on time due to evaporative cooling \cite{andresen_2010_evaporative} and the growth of diocotron modes \cite{hilsabeck_2001_finite}. The escape rate vs.\ time has a peak that shifts toward later times as $\tau$ is increased. For $\tau\geq 64\,\mathrm{s}$, the shape in (b) broadly matches the shape of Fast Scintillator count rate vs.\ time for mixing, shown in Fig.~\ref{fig:time}(d). 

We record videos \cite{hunter_2025_17434838} of the phosphor screen during extraction using a CMOS camera (Thorcam CS165MU1) exposed $50\,\mathrm{ms}$ per frame with about $5\,\mathrm{ms}$ dead time. The profiles shown in Fig.~\ref{fig:ext}(a) are frame sums for the entire extraction. The videos show that the average radius of extracted particles decreases in time, reaching a minimum when roughly half of the particles have escaped. Afterwards more violent instabilities develop, sometimes ejecting particles at $r\gg\lambda_\mathrm{D}$. 

On average, $r$ is lower for slower extraction. Extracting slowly gives particles more time to refill the depleted center. The refilling rate could be proportional to the growth rate of the fastest-growing diocotron mode. For sufficiently slow extraction we expect $r<2\lambda_\mathrm{D}$. Note that $\lambda_\mathrm{D}$ increases as particles escape and $\rho$ falls. Using a numerical solver (see Appendix~\ref{sec:appx_plasma}), we estimate $\lambda_\mathrm{D} = (0.32\pm 0.03)\,\mathrm{mm}$ at the start of mixing (Section~\ref{sec:time}). 

The relation $r<2\lambda_\mathrm{D}$ might explain why the count rate at the $\overline{\mathrm{H}}$ Beam Detector in Fig.~\ref{fig:time}(d) peaks when $T_\mathrm{p}$ is minimum. But some of the videos show an intense, low $r$ feature forming late in the extraction, which we cannot link to $T_\mathrm{p}$ so easily. Either way, it is clear why the $\overline{\mathrm{p}}$ in the DS well, shown in Fig.~\ref{fig:time}(a), have lower average radius than the ones in the US well. The DS well collects $\overline{\mathrm{p}}$ extracted from the US well, most of which escape at $r<2\lambda_\mathrm{D}$. The DS image series may be the best indication of the actual $\overline{\mathrm{p}}$ extraction radius during mixing. 

\section{Mixing duration and positron heating}
\label{sec:rate}

The results of the last section apply to $\overline{\mathrm{H}}$ production if we ignore possible effects caused by rebounding $\overline{\mathrm{p}}$. During mixing, the $\overline{\mathrm{p}}$ plasma is simply a source of $\overline{\mathrm{p}}$, most of which escape as $\overline{\mathrm{H}}$ after bouncing a few thousand times through the $\mathrm{e^+}$ plasma \cite{robicheaux_2004_simulations}. Thus, changing the mixing duration is equivalent to changing the extraction time, so we refer to both as $\tau$. Greater $\tau$ should make $\overline{\mathrm{p}}$ enter the $\mathrm{e^+}$ plasma at lower $r$ so that more $\overline{\mathrm{H}}$ atoms emerge from the trap as a beam (see previous section). For the $\mathrm{e^+}$ plasma itself, the effects of increasing $\tau$ cannot be isolated from the mixing process. To understand these effects, we will study mixing with variable $\tau$ and independently variable $T_\mathrm{e}$.

\begin{figure}[t!]
    \centering
    \includegraphics[width=\linewidth]{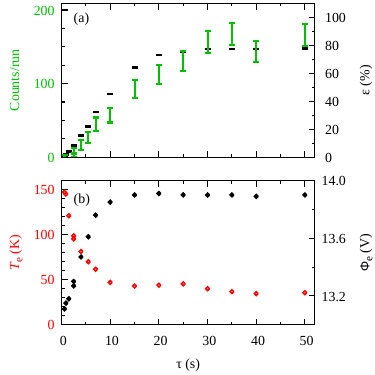}  
    \caption{Mixing for a variable duration $\tau$: (a) $\overline{\mathrm{H}}$ Beam Detector counts (green) and $\overline{\mathrm{H}}$ formation efficiency $\varepsilon$ (black), (b) $\mathrm{e^+}$ plasma temperature $T_\mathrm{e}$ (red, open circles) and space charge $\phi_\mathrm{e}$ (black, filled circles). Background $\beta$ is subtracted from each point using the formula $\beta=\beta_0+\beta_{80}\cdot\varepsilon/80\%$, where $\beta_0$ is the number of counts per run in an equivalent window without $\overline{\mathrm{p}}$ annihilations and $\beta_{80}$ is the number when $\varepsilon=80\%$ and the gate valve is closed so that no $\overline{\mathrm{H}}$ can reach the $\overline{\mathrm{H}}$ Beam Detector. Error bars in (a) are statistical, i.e., $\sqrt{\mathrm{Counts/run}}$ or the equivalent for the ratio of Fast Scintillator counts that defines $\varepsilon$.}
    \label{fig:tau}
\end{figure}

Figure~\ref{fig:tau} shows how $\tau$ affects $\overline{\mathrm{H}}$ production and $\mathrm{e^+}$ plasma properties when $1.7\times 10^6$ $\overline{\mathrm{p}}$ are mixed with a plasma containing $6.4\times 10^7\,\mathrm{e^+}$. The $\overline{\mathrm{H}}$ per $\overline{\mathrm{p}}$ efficiency $\varepsilon$, measured with the Fast Scintillator, varies in roughly the same way as the number of atoms counted by the $\overline{\mathrm{H}}$ Beam Detector. (The latter is linear in the number of $\overline{\mathrm{p}}$ used for mixing \cite{hunter_2025_measured}.) They reach a maximum for $\tau>25\,\mathrm{s}$, which is also where $\phi_\mathrm{e}$ is highest and $T_\mathrm{e}$ is lowest.

$T_\mathrm{e}$ is higher for $\tau<10\,\mathrm{s}$. The $\mathrm{e^+}$ plasma absorbs kinetic energy from the $\overline{\mathrm{p}}$ and binding energy from the $\overline{\mathrm{H}}$. The $\mathrm{e^+}$ radiate this energy away on the cyclotron cooling timescale: $T_\mathrm{e}$ falls exponentially toward its steady-state value at a rate $\Gamma^{-1}\approx 1.2\,\mathrm{s}$ for $\mathrm{e^+}$ at $z=-0.22\,\mathrm{m}$ \cite{hunter_2025_best}. The amount of energy to be absorbed scales with $\Delta N_\mathrm{p}$, while the rate at which energy is lost is approximately $\Gamma N_\mathrm{e} k_\mathrm{B}T_\mathrm{e}$. If $T_\mathrm{e}$ is to remain as low as possible, then a larger number of $\overline{\mathrm{p}}$ requires either more mixing time or more $\mathrm{e^+}$. 

The $\overline{\mathrm{H}}$ production in Fig.~\ref{fig:tau}(a) plateaus at greater $\tau$ than the $\mathrm{e^+}$ plasma properties in panel (b). To isolate the effect of $T_e$ on the other variables, we fix $\tau=50\,\mathrm{s}$ and heat the plasma using white noise. We vary the noise power $P$ on an azimuthally sectored electrode at $z=-0.17\,\mathrm{m}$ ($0.02\,\mathrm{m}$ downstream of the DS well). The noise is fed directly to two adjacent petals and phase shifted by $180^o$ on the other two, so that each random Fourier component produces a dipole field. This only heats the $\mathrm{e^+}$ (not the $\overline{\mathrm{p}}$) in the range of $P$ considered (see below). Figure~\ref{fig:heat} shows the results. $T_\mathrm{e}$ rises with increasing $P$, while $\overline{\mathrm{H}}$ Beam Detector counts, $\varepsilon$, and $\phi_\mathrm{e}$ fall.

\begin{figure}
    \centering
    \includegraphics[width=\linewidth]{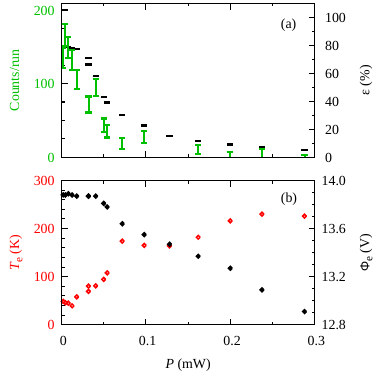}  
    \caption{Mixing while heating the $\mathrm{e^+}$ with variable noise power $P$. (a), (b), error bars, and background subtraction are the same as in Fig.~\ref{fig:tau}. $P$ is the mean-squared noise voltage applied to the electrode divided by $50\,\Omega$, the value of the termination resistor at the vacuum feedthrough.}
    \label{fig:heat}
\end{figure}

Comparing the two figures, we see that $\varepsilon$ does not depend solely on $T_\mathrm{e}$. $\varepsilon$ falls by $50\%$ for $T_\mathrm{e}\approx 60\,\mathrm{K}$ in Fig.~\ref{fig:tau}, vs.\ $T_\mathrm{e}>100\,\mathrm{K}$ in Fig.~\ref{fig:heat}. In Ref.~\cite{hunter_2025_best} we described similar experiments with only $3\times 10^5$ $\overline{\mathrm{p}}$ and $4\times 10^6\,\mathrm{e}^+$. Higher $P$ was needed to raise $T_\mathrm{e}$ of the smaller $\mathrm{e^+}$ plasma, such that $\overline{\mathrm{p}}$ were also heated by the white noise. Significant $\overline{\mathrm{p}}$ heating was observed for $P\geq 0.6\,\mathrm{mW}$. In that heating experiment, $\varepsilon$ fell by a factor of 2 for $T_\mathrm{e}>50\,\mathrm{K}$. The $\mathrm{e^+}$ plasma in Ref.~\cite{hunter_2025_best} had similar $\rho$ but four times smaller $r_0$, so a lower $T_\mathrm{e}$ would be needed in order for nascent $\overline{\mathrm{H}}$ to have on average the same number of deexcitation collisions as in a larger, hotter plasma.

$T_\mathrm{e}$ and $\phi_\mathrm{e}$ are strongly anticorrelated in our data. Removing $\mathrm{e^+}$ from the plasma reduces $\phi_\mathrm{e}$, and higher $T_\mathrm{e}$ means more $\mathrm{e^+}$ leave the plasma during mixing. To see why, define the total confinement $C$ as the difference between the local maximum of the confining potential at $z=-0.275\,\mathrm{m}$ and the local minimum at $z=-0.23\,\mathrm{m}$. It must be that $\phi_\mathrm{e}-\phi_\mathrm{p} < C$. As $C$ is reduced during mixing, particles leave the plasma, lowering $\phi_\mathrm{e}$ - $\phi_\mathrm{p}$. Whichever plasma is hotter will make up most of this difference, since collisions typically provide $k_\mathrm{B}T$ of energy to escape the well. As $T_\mathrm{e}$ increases, more $\mathrm{e^+}$ escape upstream and are wasted, which lowers $\varepsilon$ because more $\overline{\mathrm{p}}$ are left in the US well at the end of mixing. Note that higher $T_\mathrm{e}$ also reduces $\varepsilon$ because of the $T$ dependence of the 3-body recombination rate \cite{glinsky_1991_guiding}. For maximum $\varepsilon$, the $\mathrm{e^+}$ must be much colder than the $\overline{\mathrm{p}}$. How much colder depends on other factors like the self-collision rate and the rate at which the center of each plasma is refilled.

\section{Conclusion}
\label{sec:disc}

We demonstrate high efficiency antihydrogen production for slow-merge mixing of pure antiproton and positron plasmas, with $70$ to $80\%$ of the antiprotons forming stable antihydrogen. Relative to the previous state of the art \cite{ahmadi_2017_antihydrogen}, we use 30 (50) times more antiprotons (positrons) and combine them in a 60 times longer period. Longer mixing time (i) favors antihydrogen formation at lower radius, which tends to reduce the transverse velocity and increase the binding energy; (ii) prevents unnecessary positron loss, thereby increasing the number of useful antiprotons; and (iii) allows passive cyclotron radiation to keep the positrons cold as they absorb kinetic energy from the antiprotons and binding energy from the antihydrogen. These factors become increasingly important as we increase the number of antiprotons in the plasma, which is a common goal among groups that study antihydrogen \cite{bertsche_2015_physics}.

This work confirms the link between the radius at which antiprotons enter the positron plasma and the amount of beam-like antihydrogen atoms produced in slow-merge mixing. We present slow extraction as a tool to study and potentially optimize that radius. This has led to a significant increase in the number of atoms in our beam \cite{hunter_2025_measured}. 

Alternative methods could vastly increase the fraction of atoms leaving the trap as a beam \cite{jonsell_2019_formation,madsen_2021_on}. In the present method, the antiproton axial energy distribution is populated by collisions in the antiproton plasma, hence approximately thermal. It is impossible to obtain a mono-energetic beam this way. Cooling the antiprotons to lower temperature does not solve this problem, and our work suggests that such cooling may reduce the fraction of antiprotons that form antihydrogen. In the alternative methods, antiproton cooling reduces the energy spread without changing the mean energy of the beam, and, at least for Ref.~\cite{jonsell_2019_formation}, would not reduce the antihydrogen production efficiency. These methods can also be used for pulsed formation or extraction. In that case, ASACUSA could use the time of annihilation at the $\overline{\mathrm{H}}$ Beam Detector to find each atom's velocity, which is valuable for our planned spectroscopy measurements. Rejecting fast atoms increases the signal-to-noise ratio, and knowing how long each atom is in the microwave cavity allows more accurate calculation of the spin-flip probability.

\section*{Acknowledgments}
We thank Andrew Christensen for sharing his closed-form numerical plasma solver. This work was supported in part by the Istituto Nazionale di Fisica Nucleare (INFN); the Japan Society for the Promotion of Science (JSPS) KAKENHI Grants No. 19KK0075, No. 20H01930, and No. 20KK0305; the Austrian Science Fund (FWF), Grants No. P 32468 and W1252-N27; the Deutsche Forschungsgemeinschaft (DFG); and the Royal Society.

\section*{Data Availability}
The data and analysis that support the findings of this article are openly available \cite{hunter_2025_17434838}.

\bibliographystyle{tfq}
\bibliography{02_bib}

\appendix

\renewcommand{\thefigure}{A\arabic{figure}}
\setcounter{figure}{0}  
\section{Analysis of plasma properties}
\label{sec:appx_plasma}

Here we describe the models and approximations used to infer certain properties of the plasma. We first discuss the measurements from which we derive $N_\mathrm{p}$, $f$, $r_0$, and $\phi_\mathrm{p}$, then we describe the numerical solver used to validate the measurements.

\subsection{Measurements}

We count $\overline{\mathrm{p}}$ using the SiPM and MCP \cite{frederiksen_2022_counting,cripe_2024_summer}. The standard charge pick-off method using a Faraday cup cannot be applied to count $\overline{\mathrm{p}}$ because the charge collected (much more than 1 elementary charge per $\overline{\mathrm{p}}$) includes unknown contributions from annihilation products, nuclear fragments, and electrons. Instead, we will compare the experimentally observed plasma properties with self-consistent numerical solutions that satisfy the Poisson equation and Boltzmann relation in our trap. 

We use the Fast Scintillator to find $f$ in Fig.~\ref{fig:time}(b). The Fast Scintillator efficiency depends on where the $\overline{\mathrm{p}}$ annihilate. We calibrate the Fast Scintillator empirically using data from the experiments described in Sections~\ref{sec:time} and~\ref{sec:rate}. The total number of $\overline{\mathrm{p}}$ at the start of mixing ($N_p$) is constant and the number of $\overline{\mathrm{H}}$ ($N_H$) depends on the time at which mixing is interrupted (Fig.~\ref{fig:time}), $\tau$ (Fig.~\ref{fig:tau}), or $P$ (Fig.~\ref{fig:heat}). If the Fast Scintillator registers $M$ counts at mixing and $D$ counts at the null dump, we have
\begin{align}
    N_H + (N_p - N_H) &= N_p \\[8pt]
    \frac{M}{a} + \frac{D}{b} &= N_p \\[8pt]
    M \frac{b}{a} + D &= \text{constant}
    \label{eq:ab}
\end{align}
where $a$ and $b$ are the Fast Scintillator counting efficiency for mixing and null dump annihilations. The correct value of $b/a$ in Eq.~\ref{eq:ab} should yield the same number for all runs as $N_H$ varies. The value $b/a=1.41$ minimizes the standard deviation for the combined data series. To find the absolute values of $a$ and $b$, we vary the number of $\overline{\mathrm{p}}$, alternating mixing runs with runs where we use the SiPM to count the $\overline{\mathrm{p}}$ (see companion article). We obtain $a^{-1}=17$ and $b^{-1}=12$ $\overline{\mathrm{p}}$ per Fast Scintillator count at mixing and null dump, respectively. Note that the contribution from $\mathrm{e^+}$ annihilations is negligible \cite{hunter_2025_best}.

We extract $r_0$ from the measured density profile. We fit the profile to $A\cdot \exp\left[ - (r/r_0)^n \right] + B$. $A$ is an amplitude, $B$ is a constant background, $n$ controls how quickly the function decays, and $r_0$ represents the characteristic radius \cite{evans_2016_phenomenology}. We then apply a magnification factor: $r_0 = r_\mathrm{MCP}\sqrt{B_\mathrm{MCP}/B_0}$, because magnetized particles follow the diverging field lines from trap ($B_0$) to MCP ($B_\mathrm{MCP}$). For the same reason, we expect the plasma radius to vary inversely as $\sqrt{B}$ inside the trap \cite{fajans_2003_nonneutral}. That effect is small for the $\overline{\mathrm{p}}$ ($\delta B/B \approx 5\%$), and is included in our uncertainties as explained below.

\begin{figure}
    \centering
    \includegraphics[width=\linewidth]{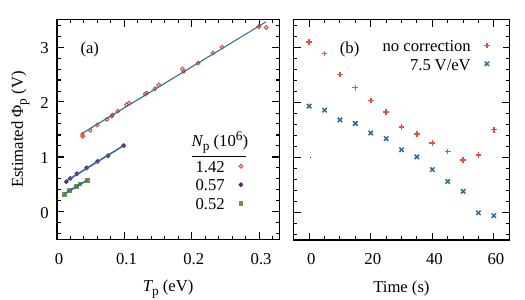}
    \caption{Calibration and measurement of $\phi_\mathrm{p}$. (a) Dependence of measured $\phi_\mathrm{p}$ on $T_\mathrm{p}$. (b) Decay of $\phi_\mathrm{p}$ during mixing. In (a), we use three different quantities of $\overline{\mathrm{p}}$ ($N_\mathrm{p}$). In (b), we show the data with and without the calibrated correction $d\phi_\mathrm{p}/dT_\mathrm{p}=7.5\,\mathrm{V/eV}$ from (a). }
    \label{fig:phivsT}
\end{figure}

We measure $\phi$ simultaneously with $T$. The first particles detected should escape when the plasma is barely confined, so the confinement potential at that time is an estimate of $\phi$. This overestimates the true value of $\phi$ by an amount proportional to $T$. The constant of proportionality should be a function of the detector sensitivity, the ramp rate, and the number of particles in the plasma \cite{hunter_2025_best}. We obtain the best estimate of $\phi$ by making a graph of measured $\phi$ vs.\ $T$, with other properties constant. Figure~\ref{fig:phivsT}(a) shows such a graph for $\overline{\mathrm{p}}$ diagnosed using the same electrodes, voltage ramps, MCP bias, and analysis routines as we use for Fig.~\ref{fig:time}(c). The $\overline{\mathrm{p}}$ are evaporatively cooled to fix $\phi_\mathrm{p}$ (and to some extent $N_\mathrm{p}$) at three different values, then heated in the DS well (see Section~\ref{sec:time}) with $0.002<P<0.2\,\mathrm{mW}$ (see Section~\ref{sec:rate}). The increase in $T_\mathrm{p}$ with $P$ is much less for lower $N_\mathrm{p}$, which incidentally corroborates our statement in Section~\ref{sec:rate} that this range of $P$ does not heat the relatively small number of $\overline{\mathrm{p}}$ in the DS well during mixing. The best-fit lines have slopes $(7.4 \pm 0.1)$, $(7.6 \pm 0.1)$, and $(7.5 \pm 0.2)\, \mathrm{V/eV}$. Figure~\ref{fig:phivsT}(b) shows estimates of $\phi_\mathrm{p}$ as a function of mixing time (Section~\ref{sec:time}), with and without applying the calibration of panel (a). The calibration brings the data much closer to the measured $f$ values shown in Fig.~\ref{fig:time}(b). It seems to slightly overcompensate for the last two points near $\phi_\mathrm{p}=0$, where the correction itself is much larger than the true value of $\phi_\mathrm{p}$.

\subsection{Numerical Solution}

The numerical solver takes as input the electrode geometry, trap potentials, $B$, $T$, $N$, and plasma rotation rate $\omega_r$ \cite{christensen_2024_exploiting}. It finds a solution with $N$ charges distributed such that $\rho(r,z)=\rho_0\,\mathrm{exp} \left[e(\Phi(r,z)-\Phi_0)/k_\mathrm{B}T \right]$. Here, $\Phi$ is the total electrostatic potential due to both the voltages applied to the electrodes and the space charge of the plasma, and $\rho_0$ and $\Phi_0$ are $\rho$ and $\Phi$ evaluated at the center of the plasma. 

We set the trap potentials to match the experimental conditions at the start of mixing and repeatedly solve for different values of $\omega_r$. For a given $T$ and $N$, this produces solutions with a range of $r_0$ values. We use these solutions for three things: First, to obtain $\rho$ and $\lambda_\mathrm{D}$, which we cannot derive from our measurements with simple formulae. Second, to verify that the properties we measure are self-consistent, i.e., correspond to a valid equilibrium state. Third, to gauge the amount of variation in the plasma properties that would still be consistent with our measurements, which we take as (an upper bound on) our systematic uncertainty.

\subsection{Uncertainties}

In Table~\ref{tab:initial} we compile measured values of $N$, $T$, $r_0$ and numerical results for $N$, $T$, $r_0$, and $\lambda_\mathrm{D}$. The statistical uncertainties for the measured values are the standard deviation in three or more repeated measurements of the same quantity. The systematic uncertainties, given with the numerical results, depend on the following factors:
\begin{enumerate}
    \item[$N_\mathrm{p}$] The detection efficiency of the MCP-phosphor-photomultiplier system is unknown. The upper bound is $100\%$. The lower bound is taken as $93\%$, as obtained in Ref.~\cite{frederiksen_2022_counting}. While we are not sure that those authors achieved high enough gain so that all events were above threshold, our measured amplitude distribution suggests that we do \cite{cripe_2024_summer}, so our detection efficiency should be at least as great as theirs.
    \item[$N_\mathrm{p}$] Pions, formed upon annihilation when $\overline{\mathrm{p}}$ hit the copper mesh at $z=-0.5\,\mathrm{m}$, may reach the MCP and emulate a $\overline{\mathrm{p}}$ signal. As the MCP is $0.4\,\mathrm{m}$ from the mesh and has a diameter $0.04\,\mathrm{m}$, the relative solid angle is less than $1\%$. We estimate that charged pions (typically two or four per $\overline{\mathrm{p}}$) coming from the $21\%$ of $\overline{\mathrm{p}}$ that annihilate on the mesh contribute less than $1\%$ of the signal.
    \item[$N_\mathrm{e}$] Secondary electrons can leave the MCP when it is used as a Faraday Cup, causing us to overestimate the positive charge of the $\mathrm{e^+}$ plasma. To characterize this effect, we vary the MCP-front bias between $-100$ and $+20\,\mathrm{V}$, measuring $N_\mathrm{e}$ simultaneously via the charge collected by the MCP and the light produced in a plastic scintillator outside the MCP chamber. The latter is proportional to the number of $\mathrm{e^+}$ that hit the MCP. A bias of $-7\,\mathrm{V}$ (or lower) maximizes the scintillator signal. However, a bias of $+25\,\mathrm{V}$ is needed to retain most of the secondary electrons \cite{knight_1995_secondary,overton_1997_measurement}. Comparing the data from scintillator and MCP at $+25$ and $-7\,\mathrm{V}$, we infer that the charge signal at $-7\,\mathrm{V}$ bias is higher than true $N_\mathrm{e}$ by about $25\%$. As elsewhere, we split the difference in our lack of certainty, taking $87\%$ of the measured value as the center and $13\%$ of the measured value as the systematic uncertainty.
    \item[$T$] The plasma temperature diagnostic \cite{eggleston_1992_parallel}, in its simplest form, assumes that $\phi$ is constant. In fact, $\phi$ decreases because of escaping charge and because the well gets longer as the confining potential is reduced, causing us to overestimate $T$ \cite{christensen_2024_exploiting}. The effect should be about $10\%$ in our case, which is comparable to the statistical variation from run to run, and consistent with the (little or no) variation in $T$ that we observe for electron plasma diagnosed from wells of different length.
    \item[$r_0$] The mirror ratio $B_\mathrm{MCP}/B_0$ cannot be measured while the trap is in place. It is taken from field maps provided by the manufacturer, which we have verified (with the trap removed) at the $1\%$ level.
    \item[$r_0$] The mirror-ratio prediction described above assumes that particles follow magnetic field lines exactly. That is an imperfect assumption, both inside the trap and between the trap and the MCP. First, the plasma changes shape as it is carried about the trap through regions of varying $B$. When it is moved from the mixing potentials, at high $B$, to the dump-diagnosis well, at low $B$, the particles should not exactly follow magnetic field lines, because that would generate large potential gradients within the plasma \cite{fajans_2003_nonneutral}. Second, particles drift transversely across field lines whenever the electric field is perpendicular to $B$, as is the case near the MCP. We quantify this effect via the discrepancy between the mirror-ratio prediction and the inferred size of the copper mesh, which appears as a shadow when the plasma is extracted through it for imaging \cite{amsler_2022_reducing}.  The discrepancy corresponds to a change in the inferred radius of about $+30\%$ for electrons, $+25\%$ for $\mathrm{e^+}$, and $-4\%$ for $\overline{\mathrm{p}}$. Thus, we obtain four predictions for the true $r_0$, depending on whether or not particles follow field lines within the trap and outside of the trap. We take the range of possible $r_0$ to be the mean of these predictions plus or minus their standard deviation.
    \item[$\lambda_\mathrm{D}$] The numerical solver assumes that the plasma can be decomposed as a set of concentric cylindrical shells of different lengths. This neglects a possible cone-shaped distortion, which should arise because $B$ varies in $z$ \cite{fajans_2003_nonneutral}. The variation of $B$, and hence $n$, is at most $10\%$ from one end of the plasma to the other. This contributes an additional $5\%$ uncertainty to $\lambda_\mathrm{D}$ (and to $r_0$), which we compound with the standard deviation described below.
\end{enumerate}
We use these uncertainties, combined in quadrature with the statistical uncertainties, to find ranges of acceptable $N$, $T$, and $r_0$, then find plasma solutions that evenly span those ranges (in $r_0$) or sit at extreme values (in $N$ and $T$). The mean and standard deviation of this set of solutions are the numerical result and the systematic uncertainty for each variable given in Table~\ref{tab:initial}.

\end{document}